\documentclass[epj]{svjour}
\usepackage{graphics}
\usepackage{epsfig}

\def\gsim{\,\lower.25ex\hbox{$\scriptstyle\sim$}\kern-1.30ex%
\raise 0.55ex\hbox{$\scriptstyle >$}\,}
\def\lsim{\,\lower.25ex\hbox{$\scriptstyle\sim$}\kern-1.30ex%
\raise 0.55ex\hbox{$\scriptstyle <$}\,}

\newcommand{\PO}{I\!\!P}

\def\Journal#1#2#3#4{{#1} {\bf #2} (#3) #4}

\def\NPB{{\em Nucl. Phys.}   {\bf B}}
\def\PLB{{\em Phys. Lett.}   {\bf B}}
\def\PRL{\em Phys. Rev. Lett.}
\def\PRD{{\em Phys. Rev.}    {\bf D}}

\def\ZPC{{\em Z. Phys.}      {\bf C}}

\def\EJC{{\em Eur. Phys. J.} {\bf C}}
\def\EJA{{\em Eur. Phys. J.} {\bf A}}

\begin{document}
\title{Pion fluctuation in deep inelastic scattering}
\author{A.~Bunyatyan\inst{1,}\inst{2} 
\thanks{\emph{e-mail:}  Armen.Bunyatyan@desy.de}
   and B.~Povh\inst{1}
\thanks{\emph{e-mail:}  B.Povh@mpi-hd.mpg.de}
}                     
\institute{Max-Planck-Institut f\"{u}r Kernphysik,
           Postfach 103980, D-69029 Heidelberg, Germany
 \and Yerevan Physics Institute, Alikhanian Brothers St.2, AM-375036
       Yerevan, Armenia
}
\abstract{
The forward neutron production in the $ep$ collisions at 300 GeV
measured by H1 and ZEUS Collaborations at DESY has been used
to estimate the total probability for proton fluctuation
into $n\pi^+$ and $p\pi^0$. The probability found is on the order of 
30\%. This number is compared with the numbers 
obtained for the probability of quark fluctuation into $\pi^+$
from several alternative DIS processes (Gottfried sum rule, 
polarized structure function) and the axial-vector coupling constant,
where the pion fluctuation is believed to play an important role. 
} 
\maketitle
%
\section{Introduction}
Measurements of the inclusive deep inelastic scattering (DIS)
have allowed to extract important information on the partonic structure
of the nucleon. 
Parton distributions have been determined and scaling violations have 
been tested to a high level of accuracy. The QCD parton model has been 
shown to be very reliable in the presence of the hard scale.

On the other hand, semi-inclusive reactions with electromagnetic probes
are less explored. Through the study of new observables characterizing
these processes we may ask more detailed questions about the hadronic 
structure. 
In semi-inclusive reactions the hadronic character of the nucleon becomes 
apparent, soft QCD becomes important.
The most prominent effects in  semi-inclusive DIS are the consequence
of the strong correlation between the sea-quark and the sea-antiquark
in the relative state with the quantum number of the pion \cite{Altarelli}. 
We will refer to this correlation, as is the custom in the hadronic physics,
as the pion fluctuation of the nucleon,
$p\to N\pi$ and $p\to \Delta\pi$.
A similar fluctuation of the proton to $K^+\Lambda$,
which is less strong than the pion fluctuation because of higher 
excitation, is the source of the strange sea in the proton and explains
the $s\bar{s}$ asymmetry in the neutrino DIS.
Also the proton fluctuation to $D\Lambda_c$ may be the source of 
intrinsic charm in proton  \cite{Brodsky,Franz}.

Two collaborations, H1 and ZEUS at HERA, have measured the production
of forward neutrons in electron-proton collisions at a center-of-mass
energy of about 300 GeV 
\cite{H1LN,zeuslninc,zeuslntraj,zeuslndstar,zeuslnjet,H1LNjets}.
 These events are successfully described within the models
based on the pion exchange mechanism: the DIS takes place on the $\pi^+$
fluctuation of the proton. Neutrons carrying more than 50\% of the proton 
initial energy produced in DIS on pions can be well
distinguished from the neutrons produced in ordinary DIS on the proton
by their energy and four-momentum distributions.

The forward neutron production with the aim to observe the pion fluctuation
has already been studied  previously in the high-energy 
$pp$-reaction at ISR in CERN \cite{ISR}. But the rescattering of proton on the 
virtual neutron is so strong that a reliable estimate of the 
probability of the $p\to\pi^+ n$ fluctuation was not possible. On the other 
hand, the rescattering of the highly virtual photon ($Q^2>10$~GeV$^2$)
on neutron is sufficiently small so that a reliable
probability for the pion fluctuation can be deduced.
Thus, the forward neutron production in DIS is probably the most direct way
to determine the probability of the pion fluctuation quantitatively. 
The pion fluctuation is very likely 
responsible for the flavour asymmetry of the quark sea in the nucleon 
as observed in the 
violation of the Gottfried sum rule. Furthermore, the reduction of 
the amount of the angular momentum carried by the quark spins in the 
nucleon is probably also, to a great extent, a consequence of a pion 
fluctuation, as we will discuss later. These two phenomena as well as the 
value of the axial current coupling constant are directly related to the 
pion fluctuation of the quarks, $u \to d\pi^+,u\pi^0$ and $d \to u\pi^-, 
d\pi^0$. The pion Compton length is larger than the nucleon radius and
the fluctuating pions overlap and interfere. They result in different 
Fock components of the nucleon with one, two and three pions.   

Therefore we felt it interesting 
to estimate the probability of the pion fluctuation
in the proton in DIS and compare it 
with the probability of the fluctuation of the quark. 
For the neutrons carrying large fraction of
the initial proton energy the energy spectra agrees well with the
pion exchange model predictions.
 At lower energies contributions from ordinary DIS are becoming important.
The H1 and ZEUS measurements were fitted by
theoretical pion fluxes from which the pion probability in the nucleon
can be deduced.
 Our present analysis investigates the uncertainty of this
probability as a consequence of the different form of the neutron 
spectra as a function of the coupling constant and the  formfactor.
The contribution of the DIS on pion at low neutron energies 
was calculated by extrapolating the neutron spectrum from energies above
50\% of the proton incoming energy to low energies and to full range of 
four-momentum transfer squared ($t$)
using three
different pion fluxes which fitted the experimental data best.

\section{Estimate of the neutron flux from DIS on pion}

 To evaluate the probability of pion fluctuation in DIS we use the 
measurements  of the leading neutron production at HERA 
performed by H1 and ZEUS Collaborations. There
the leading neutron production was studied
in different processes: in the semi-inclusive DIS process $ep\rightarrow 
enX$, in  the photoproduction of $D^*$ and in the dijet production
\cite{H1LN,zeuslninc,zeuslntraj,zeuslndstar,zeuslnjet,H1LNjets}.
The kinematic range of leading neutron measurements are
restricted by the geometrical acceptance of the forward neutron 
calorimeters to the range $x_L\gsim 0.2$ and
$\theta_n\lsim 0.8~mrad$, where $x_L$ is the neutron energy 
relative to the initial proton energy and $\theta_n$ is the
angle of the scattered neutron in the laboratory frame.

In our analysis we use only the measurements of the semi-inclusive channel
$ep\rightarrow enX$ which have the highest statistical significance
and for which one expects the least distortion providing the
measurement is done by sufficiently high $Q^2$. At low $Q^2$ the 
distortion can arise from the 
rescattering of the neutron on the extended hadronic
photon the size of which increases with decreasing $Q^2$.
The rescattering effects are expected to be small in the DIS regime, at 
sufficiently large photon virtualities $Q^2$, as is also confirmed by the
measurements \cite{zeuslndstar}.

Previous studies have shown that particle exchange models describe
the leading neutron production data both in deep inelastic scattering
at HERA and at hadroproduction  \cite{hadrexp} experiments.
In these models the transition amplitude for $p\rightarrow n$
is dominated by $\pi^+$ exchange 
\cite{Sullivan,PovhKop,Holtmann,Bishari}.
The interpretation of this process
depends on the reference frame. In the rest frame of the proton 
DIS takes place from a pion  emitted from the proton.
In the infinite momentum 
frame (the rest frame of the photon) the photon emites the pion.
Such ambiguities are resolved using light-front Fock methods 
\cite{BrodskyL}.
Let us first consider that the production of leading neutrons in DIS
at large $x_L$ proceeds entirely via $p\to n+\pi^+$ channel.
Then the cross-section for photon--proton scattering to the final
 state $nX$ takes the form

\begin{equation}
  d\sigma^{\gamma^* p\rightarrow nX}=f_{\pi^+/p}(x_L,t)\cdot 
  d\sigma^{\gamma^*\pi^+\rightarrow X},
\end{equation}

\noindent
where $f_{\pi^+/p}(x_L,t)$ is the pion flux associated with the beam 
proton and $d\sigma^{\gamma^*\pi^+\rightarrow X}$
denotes  the cross-section for the hard photon--pion interaction. 
The general form of pion flux is given by the expression

\begin{equation}
 f_{\pi^+/p}(x_L,t)=\frac{1}{2\pi}\frac{g^2_{p\pi n}}{4\pi}
(1-x_L)^{1-2\alpha(t)}\frac{-t}{(m_\pi^2-t)^2}
  |G(t)|^2.
\end{equation}

\noindent 
Here  $m_\pi$ is the pion mass and  
$g^2_{p\pi n}/4\pi=13.7$ is the $p\pi n$ coupling constant,
known from phenomenological analyses of low-energy data~\cite{Tim91}.
$G$ is a formfactor which accounts for off-mass-shell corrections, 
normalized to be unity at the pion pole.
For the light-cone approach $G$ becomes also dependent on $x_L$.
The probability of pion fluctuation depends thus on $x_L$ and $t$.
Our final result, called $<n\pi^+|p>^2$, corresponds to the 
probability integrated over $x_L$ and $t$ \cite{PIRNER,Strikman}.

\begin{figure}[h]
\centering
\epsfig{file=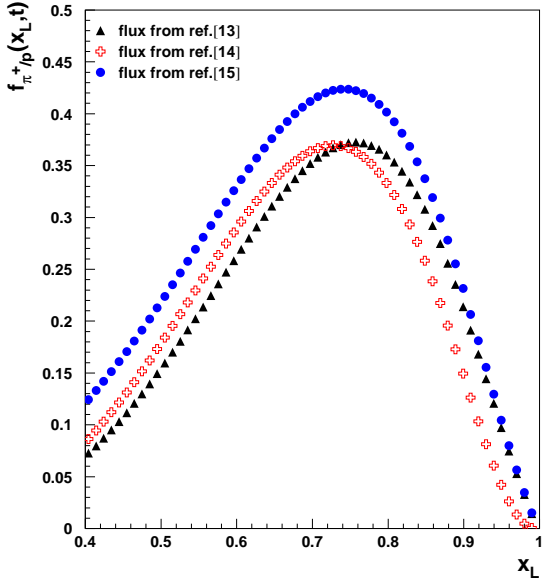,width=82mm}
  \caption{The pion flux as function of $x_L$ for different shapes
  of the formfactor integrated over $t$ in the range determined
  by the angular acceptance of the forward neutron calorimeter.}
\label{piflux}
\end{figure}

Several functional forms of pion flux can be found in the literature. 
These expressions differ in the assumptions for the
formfactor $G$ 
and in the assumption of reggeization of the pion exchange 
(e.g. the value of $\alpha(t)$).
In Fig.\ref{piflux} we show different predictions
for the shape of the neutron energy distribution.
The difference in the normalization should not affect our estimate
as our analysis depends only on the ratios of the phase spaces.
The extrapolation to the low energies and full $t$ range depends strongly 
on the choice of the flux.

We quote results using the pion fluxes of \cite{PovhKop,Holtmann,Bishari} 
which are also chosen by the experiments
\cite{H1LN,zeuslninc,zeuslntraj,zeuslndstar,zeuslnjet,H1LNjets}.

In ZEUS analysis \cite{zeuslndstar}   
the fraction of leading neutron events was measured in the
kinematic range  $Q^2>4~\rm GeV^2$, and it 
was found to be $8.0\pm 0.5\%$ for $x_L>0.2$ or $5.8\pm 0.3\%$ for 
$x_L>0.49$.
A similar result is obtained in the H1 measurements.
The result is however published only in the PhD thesis 
\cite{metlica}, quoting 
the fraction of leading neutron
events in DIS  to be $(7.9 ^{+2.0}_{-1.5})\%$ for $x_L>0.49$, 
$|t|<0.5~\rm GeV^2$.
In this analysis the inclination of proton beam with respect to the nominal
value (beam tilt) was not properly taken into account.
 Considering the increase
of geometrical acceptance for the neutrons by about 20\% due to the
observed proton beam tilt as was done for  ref.\cite{H1LNjets} gives a 
corrected value of
about $(6.3 \pm 1.5) \%$, which is in agreement with ZEUS results.
Because of the larger errors of the value obtained by H1, 
only the ZEUS result will be used in  further analysis.

Using the measured value of $(5.8 \pm 0.3)\%$ 
we estimate the full contribution of pion exchange to the DIS
by extrapolating the measurement from the measured kinematical
range to the full phase space.
 The result of this extrapolation is
 $(9.5\pm 0.5) \%$ using the flux from \cite{PovhKop},
 $(10.3\pm 0.5) \%$ for flux \cite{Holtmann} and
 $(14.6\pm 0.7) \%$ for flux \cite{Bishari}.

In this procedure we assumed that all events with leading neutrons
in the kinematic range $x_L>0.49$ originate from pion exchange
via $p\to n+\pi^+$ channel.
However, there are also other processes which can contribute.
As was estimated in \cite{PovhKop}
in the kinematic range of the measurements, the leading neutron 
production due to $\rho$ and $a_2$ exchanges, to pomeron 
exchange, or to resonance decays, is about 10\% of 
that due to pion exchange. 
The proton transition to $\Delta+\pi$ may also contribute 
to the production of leading neutrons.
Measurements of $p\to n$ and $p\to \Delta^{++}$ reactions at
Fermilab \cite{Fermilabdelta} indicate that contribution of
$\Delta$ channel to forward neutron production is
less than $6\%$.
This observation agrees with theoretical estimate of 
$\Delta + \pi$ contribution to be 
one third to two third of the $N+\pi$ probability \cite{Thomas},
if we take into account all channels which can contribute to
forward neutron production and the acceptance limitation due
to larger angular spread and lower energies of neutrons
produced from $\Delta$ decay.

Another effect which one has to take into account is the
absorption or rescattering.
It was pointed out by D'Alesio and Pirner \cite{PIRNER} that 
also for the forward neutron production in $\gamma p\to nX$ reaction
even for $Q^2>10$ GeV$^2$ the rescattering of the virtual gamma is 
not negligible. 
For the $x_L>0.5$ the average reduction of the cross-section
(e.g. increase of probability) is about 15\%.

Taking these effects into account the final values for the
corrected fraction of the DIS on the pion compared to the
total DIS cross-section is 
$(9.9 \pm 0.5)\%$ for the flux from \cite{PovhKop},
$(10.8 \pm 0.5)\%$ for the flux from \cite{Holtmann}   
and $(15.2 \pm 0.7)\%$ for the flux from \cite{Bishari}.
The big discrepancy between the results calculated using fluxes 
from \cite{PovhKop,Holtmann} 
and flux from \cite{Bishari} is the consequence of the fact that
in the flux \cite{Bishari} a constant formfactor was assumed.
The discrepancy of 30\% between the fluxes comes from the contribution
to the extrapolated cross-section from the $x_L<0.49$ region.

\begin{figure}[t]
\centering
\epsfig{file=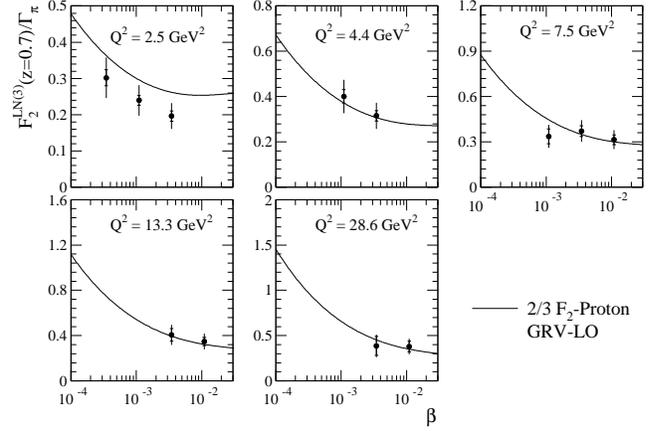,
width=90mm,bbllx=1pt,bblly=1pt,bburx=540pt,bbury=380pt,clip= }
\caption{The H1 measurement of $F_2^{LN}/\Gamma_\pi$ plotted as a function of 
       $\beta \equiv x_\pi$ for fixed 
       values of $Q^2$, where $\Gamma_\pi$ is the integrated
       pion flux factor. $F_2^{LN}/\Gamma_\pi$ can be interpreted as
       being the pion structure function $F_2^{\pi}$. 
       The data are compared
       to the GRV-LO \cite{grvp} parameterization of $F_2$ of the proton
       scaled by 2/3.}
  \label{f2pi}
\end{figure}        
\begin{figure}[h]
\centering
\epsfig{file=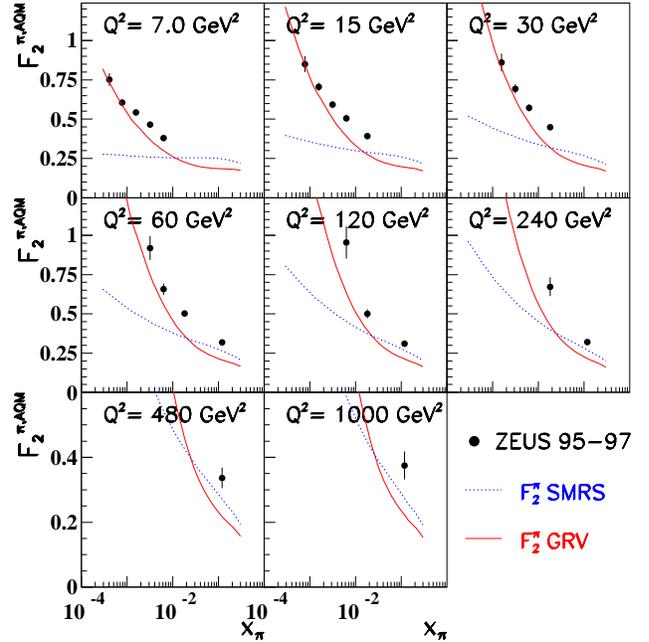,width=82mm}
\caption{The ZEUS measurement of $F_2^\pi$ in the range 
         $0.64<x_L<0.82$ as a function of $x_\pi$ 
         for  fixed $Q^2$. 
         The data are compared to the GRV \cite{grv} and 
         SMRS \cite{smrs} parameterizations of $F_2^{\pi}$.}
  \label{f2pizeus}
\end{figure}
\section{Probability of pion fluctuation observed in DIS}

To estimate the full probability of the $p\to \pi^+ n$ fluctuation,
one needs to know the relative cross-section of DIS on pion
compared to DIS on proton. 
For $x_L \sim 1$ the pion is close to mass shell and one expects the
ratio of pion cross-section to total cross-section to be 2/3.
The ratio of $F_2^{\pi}$, measured in H1 in the range $0.7<x_L<0.9$, to $F_2$
is consistent with the ratio of 2/3 \cite{H1LN,Nunneman}
(Fig.\ref{f2pi}).
In this measurement the flux from \cite{PovhKop} was used for
normalization.
 In the ZEUS analysis \cite{zeuslninc} two different normalizations  
 of flux factor were considered.
 The first method gives  ratio of $F_2^{\pi}$ to $F_2$ 
 equal to 0.361. Here the normalization was done using the
 effective flux which was taken from hadronic interactions
 \cite{Bishari}. This result contradicts the expectation that
 for $x_L=1$ the ratio is 2/3.   
 The second method used by ZEUS assumes the ratio of $\gamma^*\pi$
 to $\gamma^*p$ cross-sections to be 2/3 at $x_L= 1$.
This result is in good agreement with the GRV parameterization \cite{grv}
  of $F_2^{\pi}$ (Fig.\ref{f2pizeus}).
 The difference in the center-of-mass energies of $\gamma^* \pi$ and 
$\gamma^* p$  is also taken into account as 

\begin{equation}
\frac{\sigma^{\gamma^*\pi}(s(1-x_L))}{\sigma^{\gamma^*p}(s)}
=\frac{2}{3}(1-x_L)^{\alpha_{\PO}(0)-1}=0.6.
\end{equation}

The probability of the $\pi^+$ fluctuation of proton is then
$(16.5 \pm 0.8)\%$ for the flux from \cite{PovhKop},
$(18.0 \pm 0.9)\%$ for \cite{Holtmann} and
$(25.3 \pm 1.2)\%$ for \cite{Bishari}.
Using the isospin symmetry the fluctuation of $p\rightarrow  p+\pi^0$
is half of these numbers.
The full probability of the pion fluctuation of the proton is then
$(27.0 \pm 1.2)\%$ for the flux from \cite{PovhKop},
$(24.8 \pm 1.3)\%$ for \cite{Holtmann} and
$(38.0 \pm 1.8)\%$ for \cite{Bishari}.

 We note that the errors are entirely experimental errors.
 The systematic errors of our evaluation are reflected in the spread of 
the numbers for probabilities assuming different forms of pion fluxes.

\section{Discussion}

The pion fluctuation of the nucleon in the quark model has been
so far investigated in inclusive deep inelastic scattering:
Gottfried sum rule, spin structure functions.
From these experiments the
parameter which defines the probablity of quark
fluctuation into pion, $a=\langle d\pi^+\vert u\rangle^2$,
can be extracted.

The importance of the pion fluctuation in DIS was first realized
in the violation of Gottfried sum rule (GSR) \cite{NMC,eichten}.
The GSR expressed in parton distribution functions separated in
valence quarks ($u_v,d_v$) and sea quarks ($\bar u, \bar d$) reads 

\begin{eqnarray}
I_G(0,1;Q^2)=&&\frac{1}{3}\int_0^1 dx [u_v(x,Q^2)-d_v(x,Q^2)] \nonumber 
\\
            &-&\frac{2}{3}\int_0^1 dx [\bar{u}(x,Q^2)-\bar{d}(x,Q^2)].
\label{gott}
\end{eqnarray} 
For the symmetric sea 
\begin{equation}
I_G(0,1,Q^2)=\frac{1}{3}
\end{equation}
is obtained by noting that the first term in (\ref{gott}) becomes simply
the difference between the number of up and down valence quarks in a proton,
times 1/3. 

The second integral of (\ref{gott}) is straitforward to evaluate, if one 
assumes that the entire sea asymmetry comes from the pion fluctuation
\begin{equation}
\int_0^1 dx[\bar{u}(x,Q^2)-\bar{d}(x,Q^2)]=a(Q^2).
\label{gott1}
\end{equation}

\noindent and

\begin{equation}
I_G=\frac{1}{3}-\frac{2}{3}a.
\label{gott2}
\end{equation}

The first published results of the GSR by the New Muon 
Collaboration \cite{NMC} quotes

\begin{equation}
  I_G(0,1;Q_0^2=4~\rm GeV^2)=0.240\pm 0.016\,
\label{gott3}
\end{equation}

\noindent
This number has been afterwards corrected for the shadowing in the
deuteron measurement \cite{Antje} resulting to be 
$I_G=0.216\pm 0.024 (exp.)\pm0.009(shad.)$.
From this number using (\ref{gott2}) the probability for the pion 
fluctuation $a=0.18\pm0.05$.  
The fluctuation of the nucleon into $\Delta + \pi$ also contributes
to the correction to GSR with the same sign as $N\pi$ fluctuation.

The model used by Eichten \it et al. \rm \cite{eichten} to calculate the 
violation of the
GSR we can consider as the lowest order of the 
flavour SU(2)-chiral model of the nucleon. 
Within this model Pirner \cite{Pirner2} gives an expression for
the integral of the spin structure function as a function of $a$
as well as the expression for the axial-vector coupling constant
$g_A$.

The integral of the spin structure function $I_p$ can be written as
\begin{equation}
  I_p=\frac{5}{18}(1-2a). 
  \label{spinsf}
\end{equation}
\noindent
$I_p$ was  measured by EMC \cite{EMC} as
$I_p=0.126 \pm 0.010 (stat.) \pm 0.015 (syst.)$,
from which the  number $a$ can be obtained as
$a=0.273 \pm 0.030$.

The next best experimentally measured spin structure function is for
the deuterium. It was measured by E143 \cite{E143} as 
$I_d=0.042 \pm 0.005$.
The number $a$ is obtained from
\begin{equation}
I_D=\frac {5}{18} (1-3a)
\end{equation}
as $0.283\pm 0.040$.

We further comment on an alternative estimate of $I_P$ and $I_D$.
In order to avoid the explicite writing-down of the
baryon wave functions one assumes the validity of the flavour SU(3) 
for the hyperon octet. In the SU(3) model the semi-leptonic weak
decays can be well fitted by two parameters \cite{cabibbo}, the 
irreducible matrix elements of the SU(3) representation, F and D. 
The flavour SU(3) model is liked 
by theorists as it avoids writing explicitly the wave functions, but 
assures their proper antisymmetrisation. It may be of some disadvantage when 
applied to the spin structure functions. The axial-vector coupling constant
expressed in the model is $g_A=F+D$, and is fitted to the experimentally 
measured value $g_A=1.267\pm 0.003$.
In the SU(3) model the spin structure function depends also
on the ratio F/D. This ratio is determined 
by optimizing the fit of the semi-leptonic decays of the hyperons,
all of them having $\Delta S=1$, and takes into account possible
kaon fluctuation of the proton.
The SU(3) prediction for the integrals $I_P$ and $I_D$
using the best known values of F and D from Cabibbo \it et al. \rm 
\cite{cabibbo}
 corresponds in the SU(2)-chiral model to $a\approx 0.19$,
supporting our assumption that the main contribution to the spin
reduction comes from the pion fluctuation.
 
The axial current coupling constant may give an independent information 
on the probability of the pion fluctuation in the proton ground state.
It was early realized that the experimental value of 
the axial-vector current coupling constant of the weak decays $g_A$
can be reproduced if the pion coupling to the quarks is also taken into
account in the axial-vector current (partially conserved axial-vector current)
\cite{Adler,Weisberger}.
However, the estimate of the pion contribution is srongly model dependent.
In the SU(2) chiral model \cite{Pirner2}
\begin{equation} 
  g_A=(1-a)\frac{5}{3}. 
\label{ga}
\end{equation}

\noindent
From the measured
value of $g_A=1.267\pm 0.003$ and using eq.\ref{ga}
we obtain  $a=0.240 \pm 0.002$.

We want to compare our evaluation of the probability of
pion fluctuation in the proton ($P_{N\pi}$) 
with values of  $a$ obtained
from previously discussed inclusive processes.
In our evaluation we used  data which explicitly require
leading neutrons in the final state. 
The relation between $P_{N\pi}$ and $a$ is not obvious.
$a$ is the probability of pion fluctuation on the quark level, and
$P_{N\pi}$ is the probablity of pion fluctuation on the nucleon
level.
To connect these quantities one would need
the knowledge of the microscopic wave function of the nucleon.

\begin{figure}[h]
\centering
\epsfig{file=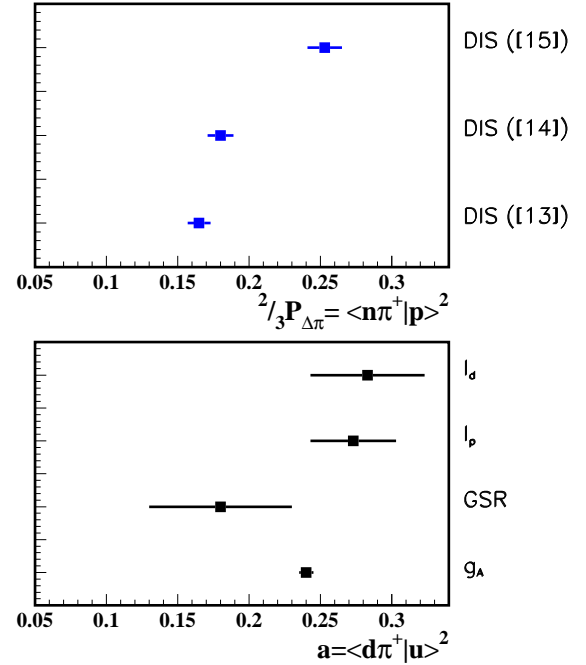,width=90mm}
\caption{Comparison of the values of the $p\to n\pi^+$ fluctuation
from DIS with $a$ from 
different experiments. DIS (\cite{PovhKop},\cite{Holtmann},\cite{Bishari}) 
are our values of $2/3P_{N\pi}$ probability 
using pion fluxes from \cite{PovhKop}, \cite{Holtmann}
and \cite{Bishari}, respectively; $I_d$ and $I_p$
are values of $a$ ($u\rightarrow d+\pi^+$ fluctuation) obtained from 
integrals of polarized structure functions
on deuteron and proton; GSR is the value of $a$ obtained from the 
Gottfried sum rule; $g_A$ is the value of $a$ from the axial-vector 
coupling.
}
  \label{a}
\end{figure}

In Fig.\ref{a}
our values of the probability of the proton fluctuation into $n+\pi^+$
(which is $2/3P_{N\pi}$) are shown together with 
the values of $a$, the $u$-quark fluctuation in $d+\pi^+$, from 
different processes.
In view of the very crude approximation of pion fluctuation being the 
ground state and ignoring the $Q^2$ dependences,
the agreement 
between the different values should be considered as reasonably good.
The observed similarity of the values of $a$ and $P_{N\pi}$
demonstrated clearly that $a$ is not an additive quantity 
\cite{Nikolaev,Strikman}, 
the pion cloud from different quarks overlap and interfere.
It is also another demonstration of the importance of 
soft interactions in semi-inclusive processes.

\section*{Acknowledgments} We wish to thank B.~Kopeliovich, H.J.~Pirner 
and W.~Weise for valuable discussions and useful remarks.



\begin{thebibliography}{99}

\bibitem{Altarelli}
G.~Altarelli, N.~Cabibbo, L.~Maiani and R.~Petronzio,
\Journal{\NPB}{69}{1974}{531}.

\bibitem{Brodsky}
S.~J.~Brodsky, C.~Peterson and N.~Sakai,
\Journal{\PRD}{23}{1981}{2745}.

\bibitem{Franz}
M.~Franz, M.~V.~Polyakov and K.~Goeke,
\Journal{\PRD}{62}{2000}{074024} [hep-ph//0002240].

\bibitem{H1LN}
C. Adloff {\it et al.}  [H1 Collaboration], \Journal{\EJC}{6}{1999}{587}
   [hep-ex/9811013].

\bibitem{zeuslninc}
S. Chekanov {\it et al.} [ZEUS Collaboration], 
\Journal{\NPB}{637}{2002}{3}
[hep-ex/0205076].

\bibitem{zeuslntraj}
S. Chekanov {\it et al.} [ZEUS Collaboration], 
\Journal{\PLB}{610}{2005}{199} [hep-ex/0404002].

\bibitem{zeuslndstar}
S. Chekanov {\it et al.} [ZEUS Collaboration], 
\Journal{\PLB}{590}{2004}{143} [hep-ex/0401017].

\bibitem{zeuslnjet}
S. Chekanov {\it et al.} [ZEUS Collaboration], 
                   \Journal{\NPB}{596}{2001}{3} [hep-ex/0010019].

\bibitem{H1LNjets}
A. Aktas {\it et al.}  [H1 Collaboration], 
    \Journal{\EJC}{41}{2005}{273} [hep-ex/0501074].

\bibitem{ISR}
J.~Engler {\it et al.}, \Journal{\NPB}{84}{1975}{70};\\
W.~Flauger and F.~M\"onnig,  \Journal{\NPB}{109}{1976}{347}.


\bibitem{hadrexp} B.~Robinson {\it et al.} \Journal{\PRL}{34}{1975}{1475}.

\bibitem{Sullivan}
J.D.~Sullivan, \Journal{\PRD}{5}{1972}{1732}.

\bibitem{PovhKop}
B.~Kopeliovich, B.~Povh and I.~Potashnikova, \Journal{\ZPC}{73}{1996}{125}
   [hep-ph/9601291].

\bibitem{Holtmann}
H.~Holtmann {\it et al.}, \Journal{\PLB}{338}{1994}{363}.

\bibitem{Bishari} M.~Bishari \Journal{\PLB}{38}{1972}{510}.

\bibitem{BrodskyL}
S.~J.~Brodsky,
 \it Nucl.~Phys.~Proc.~Suppl.\rm \bf \rm 90 (2000) 3
  [hep-ph/0009229].

\bibitem{Tim91}
R.G.E.~Timmermans, Th.A.~Rijken and J.J.~de Swart, 
        \Journal{\PRL}{67}{1991}{1074}.

\bibitem{PIRNER}
U.~D'Alesio and  H.J.~Pirner, \Journal{\EJA}{7}{2000}{109} [hep-ph/9806321].

\bibitem{Strikman}
W.~Koepf, L.~Frankfurt and M.~Strikman, \Journal{\PRD}{53}{1996}{2586}
                       [hep-ph/9507218].


\bibitem{metlica}
F.~Metlica, PhD thesis, University of Heidelberg, 1998.

\bibitem{Fermilabdelta}
J. Erwin {\it el al.,} \Journal{\PRL}{35}{1975}{980};
P.D. Higgins {\it el al.,} \Journal{\PRD}{19}{1979}{731};
S.J. Barish {\it el al.,} \Journal{\PRD}{12}{1975}{1260};
F.T. Dao {\it el al.,} \Journal{\PRL}{30}{1973}{34}.

\bibitem{Thomas}
A.~W.~Thomas and C.~Boros, \Journal{\EJC}{9}{1999}{267} [hep-ph/9812264].

\bibitem{grvp}
M.~Gl\"uck, E.~Reya and A.~Vogt, \Journal{\ZPC}{53}{1992}{127}.

\bibitem{Nunneman}
T.~Nunneman, PhD thesis, University of Heidelberg, 1998.

\bibitem{grv}
M.~Gl\"uck, E.~Reya and A.~Vogt, \Journal{\ZPC}{53}{1992}{651}.

\bibitem{smrs}
P.J.~Sutton, {\it et al.} \Journal{\PRD}{45}{1992}{2349}.
  

\bibitem{NMC}
P.~Amaudruz {\it et al.} [NMC Collaboration],  \Journal{\PRL}{66}{1991}{21};\\
M.~Arneodo  {\it et al.} [NMC Collaboration],  \Journal{\PRD}{50}{1994}{1}.


\bibitem{eichten}
   E.J.Eichten, I.Hinchliffe and C.Quigg, \Journal{\PRD}{45}{1992}{2269}.

\bibitem{Antje} A. Br\"ull, Habilitation thesis, University of Heidelberg 
1995.

\bibitem{Pirner2}
H.J.~Pirner, \it Prog.~Part.~Nucl.~Phys.\rm {\bf 36}~(1996)~19.

\bibitem{EMC}
 J.~ Ashman {\it et al.} [EMC Collaboration], \Journal{\NPB}{328}{1989}{1}.

\bibitem{E143}
K.~Abe {\it et al.}, [E143 Collaboration], \Journal{\PRL}{75}{1995}{25}.

\bibitem{cabibbo}
N.~Cabibbo, E.C.~Swallow and R.~Winston,
 \it Ann.~Rev.~Nucl.~Part.~Sci. \rm {\bf 53} (2003) 39 
 [hep-ph/0307298].

\bibitem{Adler} S.~L.~Adler, \Journal{\PRL}{14}{1965}{1051}.

\bibitem{Weisberger}W.~I.~Weisberger, \Journal{\PRL}{14}{1965}{1047}.


\bibitem{Nikolaev}
N.N~Nikolaev {\it al}, \Journal{\PRD}{60}{1999}{014004}
                        [hep-ph/9812266].

\end{thebibliography}
\end{document}